# Near-forward Raman study of a phonon-polariton reinforcement regime in the Zn(Se,S) alloy

RAMI HAJJ HUSSEIN[1], OLIVIER PAGÈS[1(a)], FRANCISZEK FIRSZT[2], AGNIESZKA MARASEK[2], WOJTEK PASZKOWICZ[3], ALAIN MAILLARD[4] and LAURENT BROCH[1]

[1] *LCP-A2MC, Institut Jean Barriol, Université de Lorraine, France*
[2] *Institute of Physics, N. Copernicus University, 87-100 Toruń, Poland*
[3] *Institute of Physics, Polish Academy of Sciences, 02-668 Warsaw, Poland*
[4] *LMOPS, Université de Lorraine – Sup´elec, 2, rue Edouard Belin, 57070 Metz, France*

**Abstract** – We investigate by near-forward Raman scattering a presumed reinforcement of the (A-C,B-C)-mixed phonon-polariton of a $A_{1-x}B_xC$ zincblende alloy when entering its longitudinal optical (*LO*-)regime near the Brillouin zone centre Γ, as predicted within the formalism of the linear dielectric response. A choice system to address such issue is $ZnSe_{0.68}S_{0.32}$ due to the moderate dispersion of its refractive index in the visible range, a *sine qua non* condition to bring the phonon-polariton insight near Γ. The *LO*-regime is actually accessed by using the 633.0 nm laser excitation, testified by the strong emergence of the (Zn-Se,Zn-S)-mixed phonon-polariton at ultimately small scattering angles.

Due to the polarity of the chemical bonding in such a ionic crystal as a zincblende AB semiconductor compound, the long-wavelength (Γ-like, $q$~0) transverse optical (*TO*) phonon, corresponding to anti-phase displacement of the intercalated A-like and B-like *fcc* sublattices (mechanical character), is likely to be accompanied by a macroscopic electric field [1]. The latter is transversal to the direction of propagation, thus identical in nature to that carried by a pure electromagnetic wave, namely a photon. Now, due to the quasi vertical dispersion of a photon at the scale of the Brillouin zone, the electromagnetic character of a *TO* mode can only emerge very close to Γ. The concerned $q$ values are of the order of one per ten thousands of the Brillouin zone size [2]. At this limit the electromagnetic and mechanical characters combine, conferring on a *TO* mode the status of a so-called phonon-polariton (*PP*). For certain $q$ values the *PP* might acquire a dominant electromagnetic character, thus propagating at lightlike speeds. This stimulates interest in view of ultrafast (photon-like) signal processing at THz (phonon-like) frequencies [3].

The $\omega$ vs. $q$ dispersion of *PP*'s propagating in the bulk of various AB zincblende compounds have been abundantly studied, both experimentally and theoretically [2,4-9]. In a nutshell, it can be grasped within four asymptotic behaviors, i.e. two photon-like ones ($\omega$-related) and two phonon-like ones ($q$-related). For large $q$ values, i.e. falling within few percent of the Brillouin zone size, as routinely accessible in a conventional backscattering Raman experiment (schematically operating in a "reflection mode", see below), a transverse electric field cannot propagate at THz (phonon-like) frequencies, because the considered ($\omega,q$)-domain falls far away from the natural dispersion of a photon (quasi vertical). In such so-called $q_\infty$-regime, a *TO* mode thus reduces to a purely mechanical oscillator (abbreviated $PM-TO$ hereafter, deprived of electric field), whose frequency, noted $\omega_{TO}$, constitutes the first phonon-like asymptote, i.e. away from Γ. The frequency of the non-dispersive longitudinal optical (*LO*) mode, noted $\omega_{LO}$, larger than $\omega_{TO}$ [10], defines the second phonon-like

(a) Corresponding author
E-mail: olivier.pages@univ-lorraine.fr

asymptote, near Γ then, taking into account that the *TO* and *LO* modes are degenerate strictly at Γ [11]. Two remaining photon-like asymptotes determine limit *PP*-behaviors away from the (*PM−TO)−LO* resonance, as dictated by the static $\varepsilon_0$ (ω> $\omega_{TO}$) and high-frequency $\varepsilon_\infty$ (ω>>$\omega_{TO}$) relative dielectric constants of the crystal. The strong *PP* coupling occurs when the quasi vertical photon-like asymptotes cross the horizontal *TO* and *LO* phonon-like ones. This gives rise to an anticrossing, resulting in two distinct *PP* branches. The upper branch is phonon-like (*LO*) when $q\rightarrow 0$ and photon-like ($\omega = q\times c\times\varepsilon_\infty^{-1}$ where c represents the speed of light in vacuum) at $\omega >> \omega_{TO}$, while the lower branch is photon-like ($\omega = q\times c\times\varepsilon_0^{-1}$) at $\omega << \omega_{TO}$ and phonon-like in the $q_\infty$-regime (*PM−TO*). Note that the (*PM−TO) −LO* band is forbidden for the propagation of bulk *PP*'s, only surface *PP*'s can propagate therein.

An interesting question is how such *PP* picture modifies for a multi-oscillator system such as a $AB_{1-x}C_x$ zincblende alloy (A standing for a cation or an anion)? Bao and Liang provided a pioneering theoretical insight into the '*ω* vs. *q*' *PP*−dispersion of various $AB_{1-x}C_x$ zincblende alloys [12,13]. As a starting point they assumed a crude two-mode [1×(A−C),1×(A−C)] *PM−TO* pattern behind the *PP's*, as explained within the modified-random-element-isodisplacement (MREI) model [14]. Besides the lower and upper alloy-related branches, assimilating to those of a pure compound as described above [2,4-9], an intermediary (A−C, B−C) –mixed *PP* was predicted by Bao and Liang. The latter branch is distinct in nature from the former parent-like two in that it exhibits an overall S-like shape governed by two phonon-like asymptotes only, i.e. the higher *PM−TO* frequency in the $q_\infty$-regime, say the BC−like one, and the lower *LO* frequency near Γ, the AC-like one then [12,13]. As such, its dispersion covers the gap between the natural *A−C* and *B−C* vibration frequencies, possibly a considerable one depending on the alloy.

We have refined the above *PP*−picture at the occasion of a recent near-forward Raman study (schematically operating in a "transmission mode", see below) of the $Zn_{0.67}Be_{0.33}Se$ zinblende alloy characterized by a three-mode [1×(*Zn−Se*),2×(*Be−Se*)] *PM−TO* pattern in the $q_\infty$-regime. Such three-mode pattern falls beyond the scope of the MREI scheme. It was explained by introducing the phenomenological percolation model [15]. In brief, this model distinguishes between the vibrations of the short (Be-Se) bonds depending on whether their local environment is more rich of one or the other (Zn-Se or Be-Se) species (1-bond→2-mode behavior) [16]. At this occasion, two intermediary (Zn-Se,Be-Se)-mixed *PP*'s were revealed, and not only one as predicted within the MREI scheme. Each intermediary *PP* relates to a particular BeSe-like *PM−TO* in the $q_\infty$-regime, and collapses with an S-like shape onto the *LO* immediately underneath near Γ, consistently with the basic MREI trend (see above). In particular, the lower-intermediary *PP*, noted $PP^-$, attracts attention, for two reasons. First, in contrast with the upper-intermediary *PP*, noted $PP^+$, which remains confined within the upper doublet of *PM−TO*'s (BeSe-like in this case), $PP^-$ may exhibit a considerable dispersion, as discussed above within the MREI scheme. Second, as soon as entering the *PP* regime, $PP^-$ becomes much dominant over $PP^+$.

The *q*-dependence of the $PP^-$ Raman intensity constitutes *per se* an interesting issue. Based on our recent near-forward Raman study of $Zn_{0.67}Be_{0.33}Se$ limited to the early stage of the *PP* regime (a moderate $PP^-$ red-shift of ~15 cm$^{-1}$ was detected, representing less than ten percent of the total $PP^-$ dispersion), we already know that the initial *q*-induced softening of $PP^-$ goes with a progressive collapse of this mode [15]. Now, as $PP^-$ is supposed to assimilate to a *LO* near Γ [12,13,15], *a priori* showing up strong and sharp in the Raman spectrum, we anticipate that, after its initial collapse, $PP^-$ should reinforce. This points to a specific *PP* feature of an alloy, yet unexplored, neither experimentally nor theoretically.

In this work we tackle such issue both theoretically, within the formalism of the linear dielectric response, and experimentally, by applying the near-forward Raman scattering to the $ZnSe_{0.68}S_{0.32}$ zincblende alloy. Due to large optical gaps of ZnSe (2.7 eV) and ZnS (3.6 eV) [17], $ZnSe_{0.68}S_{0.32}$ is transparent to visible laser lines, thus well-suited in view of a near-forward Raman study. Besides, the frequency gap between the *PM −TO* 's of its (Zn-Se and Zn-S) constituting bonds is narrow (235–285 cm$^{-1}$, see below), with concomitant impact on the magnitude of the $PP^-$ dispersion. This offers an opportunity to explore the $PP^-$ dispersion in a different context than was earlier done with $Zn_{0.67}Be_{0.33}Se$ [15], the latter alloy being characterized by a huge $PP^-$ dispersion (250 – 450 cm$^{-1}$). Basically, our aim is to penetrate deep into the $PP^-$ dispersion of $ZnSe_{0.68}S_{0.32}$ so as to address minimal *q* values likely to fall into the *LO*-regime of $PP^-$, near Γ.

Generally, the wavevector $\vec{q}$ of a *TO* mode (a *PP* one near Γ or the corresponding *PM −TO* one in the $q_\infty$-regime) accessible in a Raman experiment is governed by the conservation rule $\vec{q} = \vec{k_i} - \vec{k_s}$ in which $\vec{k_i}$ and $\vec{k_s}$ are the wavevectors of the incident laser beam and of the scattered light, respectively, both taken inside the crystal, forming an angle *θ*. In a standard backscattering geometry $\vec{k_i}$ and $\vec{k_s}$ are (nearly) antiparallel (*θ*~180°), so that *q* is maximum, falling deep into the $q_\infty$-regime. Minimum *q* values, i.e. those likely to address the *PP*-regime, are achieved by taking $\vec{k_i}$ and $\vec{k_s}$ (nearly) parallel (*θ*~0°), using a near-forward Raman setup. From the above conservation rule,

$q = c^{-1}\times\{n^2(\omega_i,x)\times\omega_i^2 + n^2(\omega_s,x)\times\omega_s^2$

$-2\times n(\omega_i,x)\times n(\omega_s,x)\times\omega_i\times\omega_s\times\cos\theta\}^{0.5}$ (1)

where $\omega_i$ and $\omega_s$ are the frequencies of the incident and scattered lights, respectively, and $n(\omega_i,x)$ and $n(\omega_s,x)$ the corresponding refractive indexes of the considered $AB_{1-x}C_x$ alloy. If we refer to pure ZnSe and pure ZnS, the refractive index of $ZnSe_{1-x}S_x$ is expected to decrease with the frequency in the visible range. Accordingly one cannot achieve $q$=0 (Γ) in practice. Indeed the minimum $q$ value, accessed in a perfect forward scattering experiment ($\theta\sim 0°$), i.e. $q_{min} = |n(\omega_i)\times\omega_i - n(\omega_s)\times\omega_s|$, remains finite because the difference in frequencies is augmented by the difference in refractive indexes. Optimum conditions are thus reached by minimizing the dispersion of the refractive index around the used laser excitation. The alloy composition ($ZnSe_{0.68}S_{0.32}$) as well as the laser excitation ($\omega_i$) used, were selected in this spirit, as discussed below.

The $ZnSe_{0.68}S_{0.32}$ sample considered in this work was grown from the melt as a single crystal (cylinder, ~3 mm-high and ~8mm in diameter) by using the high-pressure Bridgman method (see detail, e.g. in Ref. [18]). The near-forward and backward Raman spectra are taken along the [110]-growth axis, corresponding to a nominal (*TO*-allowed, *LO*-forbidden) geometry, after optical polishing of the opposite (110) faces to optical quality until quasi parallelism was achieved.

The selected alloy corresponds to the highest achievable S incorporation by the Bridgman method. This provides optimal conditions with respect to the dispersion of the refractive index, since the latter is larger for ZnSe than for ZnS in the visible range [19]. In fact, the wavelength (λ) dependence of the $ZnSe_{0.68}S_{0.32}$ refractive index ($n$) measured by ellipsometry in the λ-range 450 – 770 nm (not shown), can be accurately fitted to the Cauchy dispersion formula

$$n(\lambda) = X + Y\times\lambda^2\times 10^4 + Z\times\lambda^{-4}\times 10^9 \quad (2)$$

using $(X,Y,Z)$ values of ($3.1638 \mp 0.0007$, $1.2380 \mp 0.0536$, $14.0485 \mp 0.1056$), where $X$ is dimensionless, $Y$ and $Z$ are constants with units of $nm^2$ and $nm^4$, respectively, and λ is in *nm*. The minimum dispersion of the $ZnSe_{0.68}S_{0.32}$ refractive index is further achieved by shifting the Raman analysis to the less energetic end of the visible spectral range. We accordingly use the 633.0 nm line from a He-Ne laser to record the $ZnSe_{0.68}S_{0.32}$ near-forward Raman spectra, instead of the 514.5 nm and 488.0 nm Ar+ ones earlier used with $Zn_{0.67}Be_{0.33}Se$ [15], notwithstanding its inferior Raman efficiency. Along the same line, the Stokes experiment ($\omega_i>\omega_s$) was preferred against the anti-Stokes one ($\omega_i<\omega_s$), not to mention that the Stokes process is, further, more efficient.

Preliminary insight as to whether there is any chance to access experimentally the presumed *LO*-like $PP^-$ reinforcement with $ZnSe_{0.68}S_{0.32}$ in a near forward Raman scattering experiment using the 633.0 nm laser excitation is achieved by calculating the related multi-*PP* near-forward Raman cross section (*RCS*) in its $(q,\theta)$-dependence. For doing so we use the generic *RSC* expression established in Ref. [15], which we reproduce hereafter in its specific form applying to multi-*PP*'s propagating in the bulk of an alloy containing $p$ oscillators (in reference to the multi $PM-TO$' s of the $q_\infty$-regime behind the *PP*'s),

$$\mathrm{RCS}\,(\omega,q,x) \sim \mathrm{Im}\,\{-[\varepsilon_r(\omega,x)-q^2c^2\omega^{-2}]^{-1} \times [1+\sum C_p(x)L_p(\omega,x)]^2 + \sum C_p^2(x)\,\omega_p^2(x)\,L_p(\omega,x)\,S_p^{-1}(x)\,\varepsilon_{\infty,p}^{-1}\,\omega_p^{-2}\} \quad (3)$$

$\varepsilon_r(\omega, x)$ is the relative dielectric function of the $AB_{1-x}C_x$ alloy taken in its classical form, including the electronic contribution, $\varepsilon_\infty(x)$, varying linearly with $x$ between the parent values, and a summation over the $p$ ionic oscillators present in the crystal, each represented by a damped Lorentzian resonance $L_p(\omega,x) = \omega_p^2(x)\times(\omega_p^2(x)-\omega^2-j\times\gamma_p(x)\times\omega)$ at the relevant $PM-TO$ frequency $\omega_p(x)$ in the $q_\infty$-regime. The damping parameter $\gamma_p(x)$ is sample-dependent. $C_p(x)$ and $S_p(x)$ are the $p$-related Faust-Henry coefficient and oscillator strength, monitoring the Raman intensity and the $(PM-TO)-LO$ frequency gap of oscillator $p$, respectively. Both parameters scale linearly with the fraction of oscillator $p$ in the crystal. The ($C_p$, $S_p$, $\omega_p$, $\varepsilon_{\infty,p}$) values of the pure ZnSe or ZnS compound related to oscillator $p$ are (−0.7, 2.92, 254.5 $cm^{-1}$, 5.75) [15] and (−0.45 [20], 2.57 [21], 277.0 $cm^{-1}$ [22], 5.20 [22]), respectively. Remaining alloy-related input parameters are the number $p$ of oscillators *per* alloy, the corresponding $\omega_p(x)$ frequencies and the fractions $f_p(x)$ of such oscillators at a given composition $x$. Such [p, $\omega_p(x)$, $f_p(x)$] parameters refer to the $q_\infty$-regime, and are thus accessible from a pure-*TO* backscattering Raman insight or by infrared absorption.

Vinogradov *et al.* have revealed a three-mode $[1\times(Zn-Se),2\times(Zn-S)]$ $PM-TO$ pattern for $ZnSe_{1-x}S_x$ ($p=3$) in their recent exhaustive infrared study [22]. The question as to whether the Zn-S doublet falls into the scope of the 1-bond→2-mode percolation scheme or not will be debated elsewhere. For our present use we may only retain that the two Zn-S $PM-TO$'s, presently labelled as $TO_{Zn-S,1}$ and $TO_{Zn-S,2}$ show up distinctly in the backscattering Raman spectrum of $ZnSe_{0.68}S_{0.32}$, being characterized by comparable Raman intensities (a direct insight is given below). This means that the available Zn-S oscillator strength, which scales as the Zn-S bond fraction, equally shares between the two Zn-S submodes, leading to $f_{Zn-S,1}=f_{Zn-S,2}$=0.16. As for the unique Zn-Se mode, its oscillator strength directly scales as the Zn-Se bond fraction, so that $f_{Zn-Se}$=0.68. Last, the three $[TO_{Zn-Se},TO_{Zn-S,1},TO_{Zn-S,2}]$ $PM-TO$'s emerge at the $\omega_p(x)$

Raman frequencies of (~210, ~285, ~303) $cm^{-1}$, correspondingly.

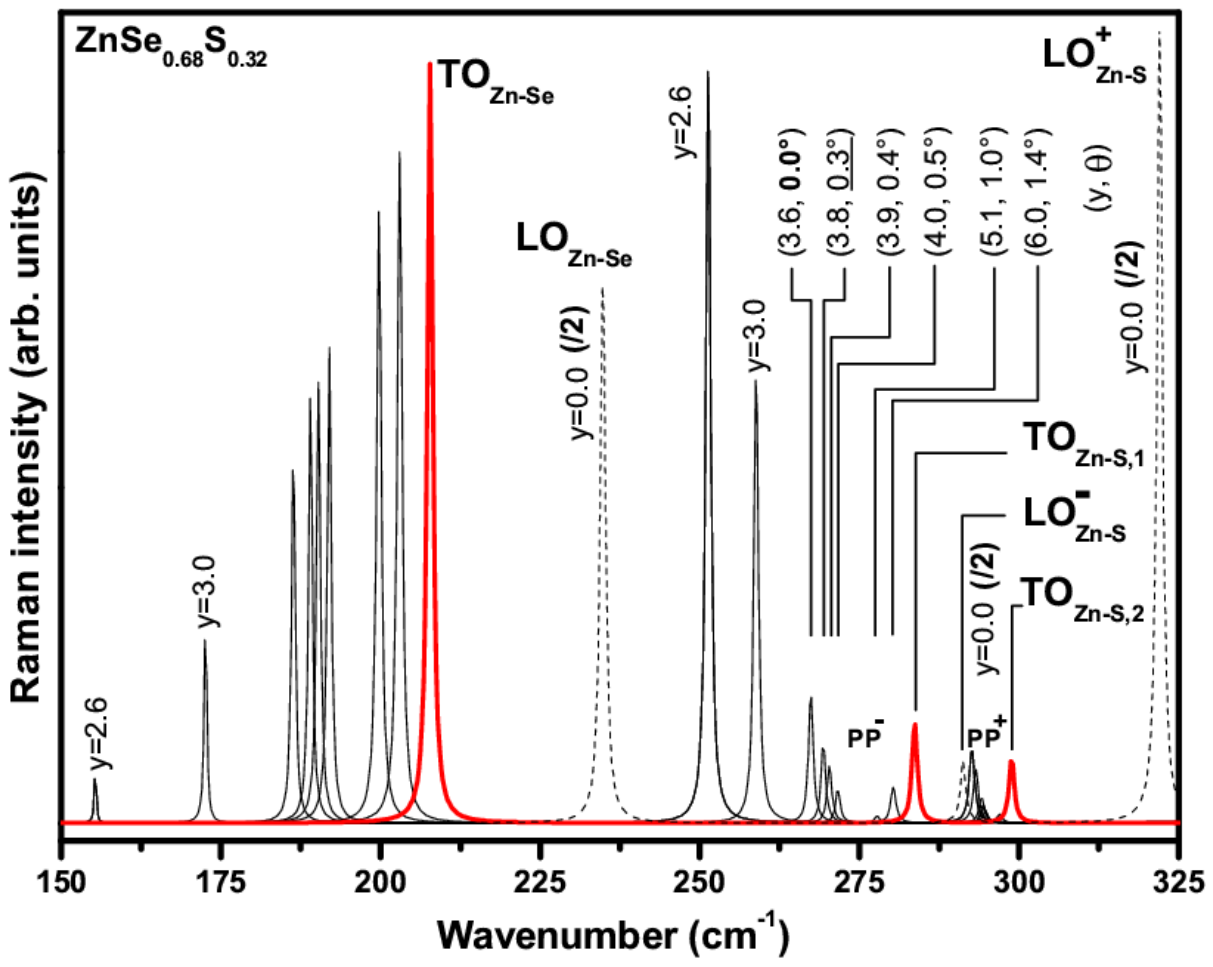

Fig. 1: (Colour on-line) Theoretical *RSC(γ)* multi-*PP* near-forward Raman lineshapes of the three-oscillator $ZnSe_{0.68}S_{0.32}$ alloy. A uniform phonon damping $\gamma_p(x)$ of 1 $cm^{-1}$ ($p$=1-3) is taken. The $\theta$ angles for the relevantpeaks addressed in a near-forward Raman scattering experiment using the 633.0 nm laser excitation are indicated. The theoretical (bold) and experimental (underlined, in reference to **Fig. 2**) $\theta$−limits are emphasized. For clarity, a simple *TO*-labeling is used for the *PM-TO*'s of the reference backscattering signal (bold-red).

The theoretical $q$-dependent near-forward multi-*PP* Raman lineshapes derived for $ZnSe_{0.68}S_{0.32}$ via **Eq. (3)** using the above $[p,\omega_p(x),f_p(x)]$ input parameters are shown in **Fig 1**, taking a uniform phonon damping of 1 $cm^{-1}$, for reasonable resolution. In fact, instead of $q$ we conveniently use the dimensionless parameter $y=q\times c\times\omega_1$ taking arbitrarily $\omega_1$, the *PM −TO* frequency of the pure ZnSe crystal in the $q_\infty$-regime, for the change of variable. As anticipated, in the theoretical *RCS*($y$) curves the initial $PP^-$ collapse at large $y$ values is relayed by a *LO*−like $PP^-$ reinforcement at moderate-to-small $y$ values. Interestingly, a moderate penetration into the $PP^-$ dispersion curve, not exceeding one third of the total dispersion (covering the $TO_{Zn-Se}$ – PM- $TO_{Zn-S,1}$ frequency gap, see **Fig 1**), seems sufficient to access the *LO*−like reinforcement regime. The crucial question is whether the latter can be accessed experimentally, or not. Basically this depends on the dispersion of the refractive index around the used laser excitation, as already discussed. For a direct insight we indicate in front of a given theoretical $PP^-$ peak in **Fig. 1**, corresponding to a certain $\omega = |\,\omega_i - \omega_s\,|$ frequency at a certain $y$ value, the relevant $\theta$ angle for the used 633.0 nm laser excitation, as directly inferred from **Eq. (1)** via **Eq. (2)**. As shown in **Fig. 1** the minimum accessible $y$ value with the latter laser line, corresponding to $\theta$ = 0°, hopefully falls into the $PP^-$ reinforcement regime. This is encouraging in view of a possible detection by near-forward Raman scattering using the 633.0 nm excitation [25].

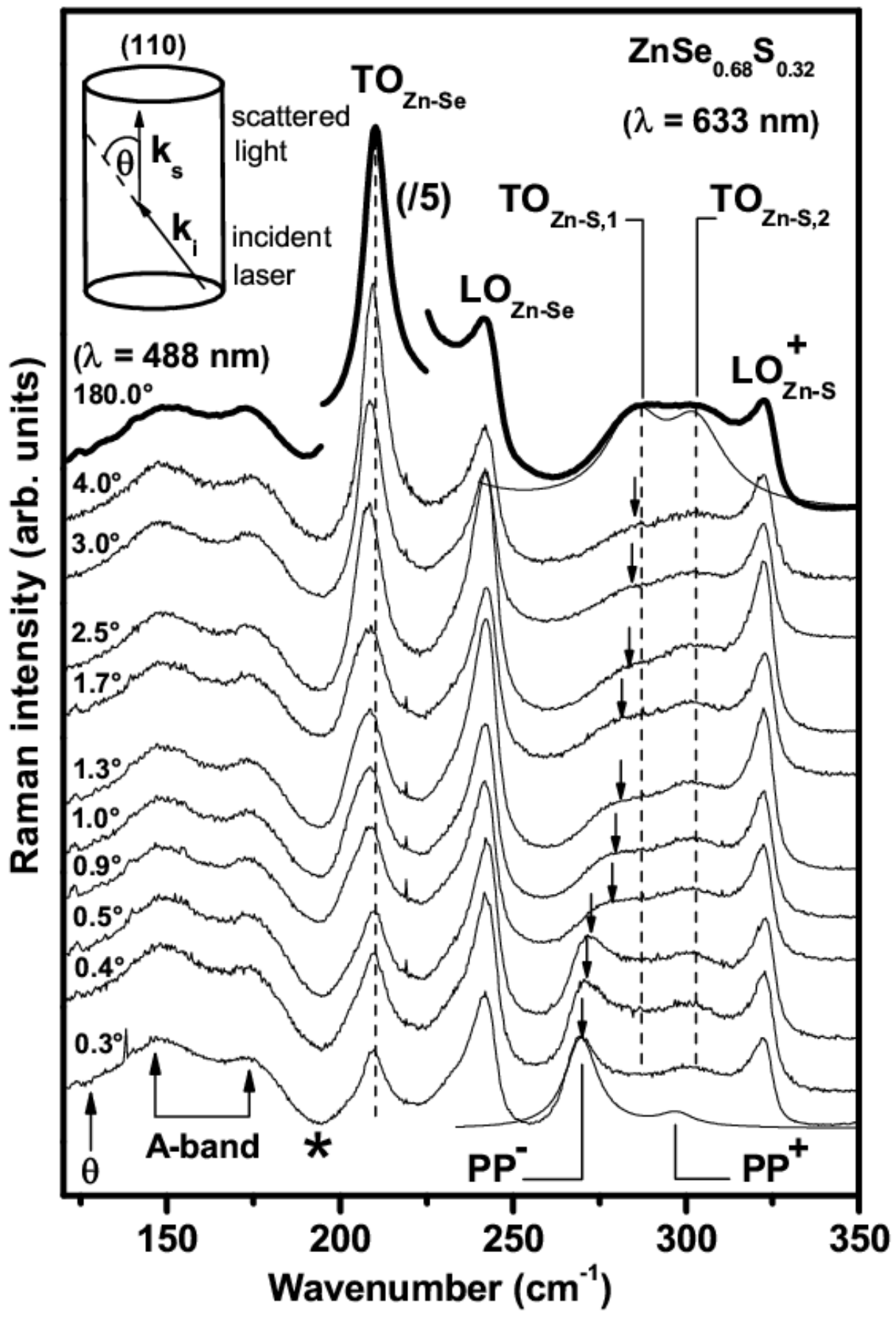

Fig. 2: Intensity-normalized $ZnSe_{0.68}S_{0.32}$ near-forward Raman spectra taken with the 633.0 nm laser excitation. The nominal/near-forward (*PP* labeling) and parasitical/backward (a simple *TO* labeling is used for the *PM-TO*'s, marked by dashed lines) Raman signals are distinguished. The $ZnSe_{0.68}S_{0.32}$ backward Raman spectrum taken with the 488.0 nm laser excitation is added (top-thick curve), for reference purpose. The theoretical (thin lines) backward ($\theta$~180°) and near-forward ($\theta$~0.3°) Zn-S Raman signals are superimposed onto the corresponding experimental ones, for comparison.

We turn to experiment. For reference purpose, we show at the top of **Fig. 2** (thick spectrum) the nominally pure-*TO* Raman spectrum taken in a backscattering geometry along the [110]-growth crystal axis using the most energetic laser line at hand, namely the 488.0 nm one. This line is more absorbed than the 633.0 nm one in our sample, which minimizes the parasitical near-forward Raman signal due to reflection of the incident laser at the back surface of the (quasi) transparent sample. Along with the nominal triplet ($TO_{Zn-Se}$, $TO_{Zn-S,1}$,$TO_{Zn-S,2}$) of *PM −TO*'s described above, several theoretically forbidden features are observed, such as the $LO_{Zn-Se}$ and $LO_{Zn-S}^{+}$ modes at ~245 and ~322 cm$^{-1}$, respectively, and also the so-called *A*–band around 165 cm$^{-1}$, reflecting the acoustical two-phonon density of states. We mention that the emergence of theoretically forbidden optic modes or acoustic bands is a common feature of alloys, due to a partial breaking of the wavector conservation rule by the alloy disorder. As for the minor $LO_{Zn-S}^{-}$ mode of the Zn-S doublet [15,16], this remains screened by the related *PM−TO*'s nearby (see **Fig. 1**). Satisfactory contour modeling of the nominal three-mode *PM−TO* pattern (top thin curve in **Fig. 2**) is achieved both in the Zn-Se (not shown) and Zn-S (thin line) spectral ranges by injecting our above selection of input parameters into the asymptotic form of **Eq. (3)** valid in the $q_{\infty}$-regime (backscattering-like), in fact reduced to its second term, taking the same $\gamma_p(x)$ damping of 24 cm$^{-1}$ for the two Zn-S sub-modes.

Now we change the Raman setup from backscattering to near-forward scattering. A representative series of near-forward Raman spectra taken at different $\theta$ angles with respect to the [110]-growth axis of the $ZnSe_{0.68}S_{0.32}$ ingot using the 633.0 nm laser line are shown in **Fig. 2** (thin spectra). The intensities, normalized to the $\theta$-insensitive A–band, are directly comparable. In each spectrum the near-forward Raman signal (*PP*−labeling in **Figs. 1** and **2**) is partially obscured by the $\theta$-insensitive backscattering one (*TO*−labeling in abbreviation of *PM−TO*) generated after reflection of the laser beam at the top surface (detector side) of the sample. As $PP^{+}$ is located within the Zn-S doublet of parasitical *PM−TO*'s (see **Fig. 1**), it remains screened at any angle, only $PP^{-}$, falling outside the doublet, is visible. Its $(q,\theta)$−dependence is discussed hereafter.

The relevant $\theta$ angle per spectrum (see **Fig. 2**) is determined by adjusting the frequency of the theoretical $PP^{-}$ peak obtained by **Eq. (3)** to the experimental frequency. As long as $\theta$ is larger than ~4.0°, the near-forward Raman signal of $ZnSe_{0.68}S_{0.32}$ remains stable, similar to the reference backscattering one ($\theta$~180°). Below this critical angle, the multi-*TO*'s enter their *PP*-regime, signed by the red-shift of $PP^{-}$ (see arrows in **Fig. 2**). Note that the minimum achievable $\theta$ angle remains finite (~0.3°), meaning that the near-forward Raman study cannot be fully developed in practice. This, we attribute to slight $\vec{k_i}$ −disorientations inside the crystal due to inherent defects in an alloy. Nevertheless, the accessible $\theta$-domain suffices to reach the *LO*-like reinforcement regime ($\theta$ :1.0°→0.3°) beyond the initial collapse regime ($\theta$ :4.0°→1.3°, see **Figs. 1** and **2**), testified by the development of $PP^{-}$ into a sharp and intense feature at near-normal incidences. Fair contour modeling of the peak $PP^{-}$ at the minimum achievable $\theta$ angle (~0.3°) is obtained by using the same input parameters as specified above, taking a reduced damping of 12cm$^{-1}$. We emphasize that the backscattering (*PM−TO*'s abbreviated *TO*'s, upper thin curve in **Fig. 2**) and near-forward ($PP^{-}$, lower thin curve) theoretical signals are directly comparable. Incidentally, we have checked that a similar near-forward Raman study with $Zn_{0.67}Be_{0.33}Se$ (see Ref. [15]) using the 633.0 nm laser excitation falls short of engaging the $PP^{-}$ reinforcement regime, due to an unfavorable dispersion of the refractive index.

Last, we discuss briefly the Zn-Se signal. At first sight it remains $\theta$-insensitive, pinned by a Fano interference with the *A*–band nearby (a characteristic antiresonance is marked by an asterisk in **Fig. 2**), as earlier discussed for $Zn_{0.67}Be_{0.33}Se$ [15]. The dramatic weakening of the near-forward Zn-Se signal with respect to the backscattering one (divided by ~5, see **Fig. 2**) is attributed to a reinforcement of the Fano interference that occurs when the Zn-Se mode engages its *PP* regime and starts to red-shift towards the $\theta$-insensitive *A*–band. Now, a careful examination reveals that the Zn-Se line develops an increasing asymmetry on its low-frequency side when $\theta$ reduces, eventually disappearing at near-normal incidences. In the process, the Raman intensity reduces dramatically. A comparison with the reference *A*–band is explicit with this respect. The discussion of the appearance and then disappearance of such asymmetrical broadening under $\theta$ reduction, presumably related to some fine structuring of the ZnSe-like *PP*, falls beyond the scope of this work.

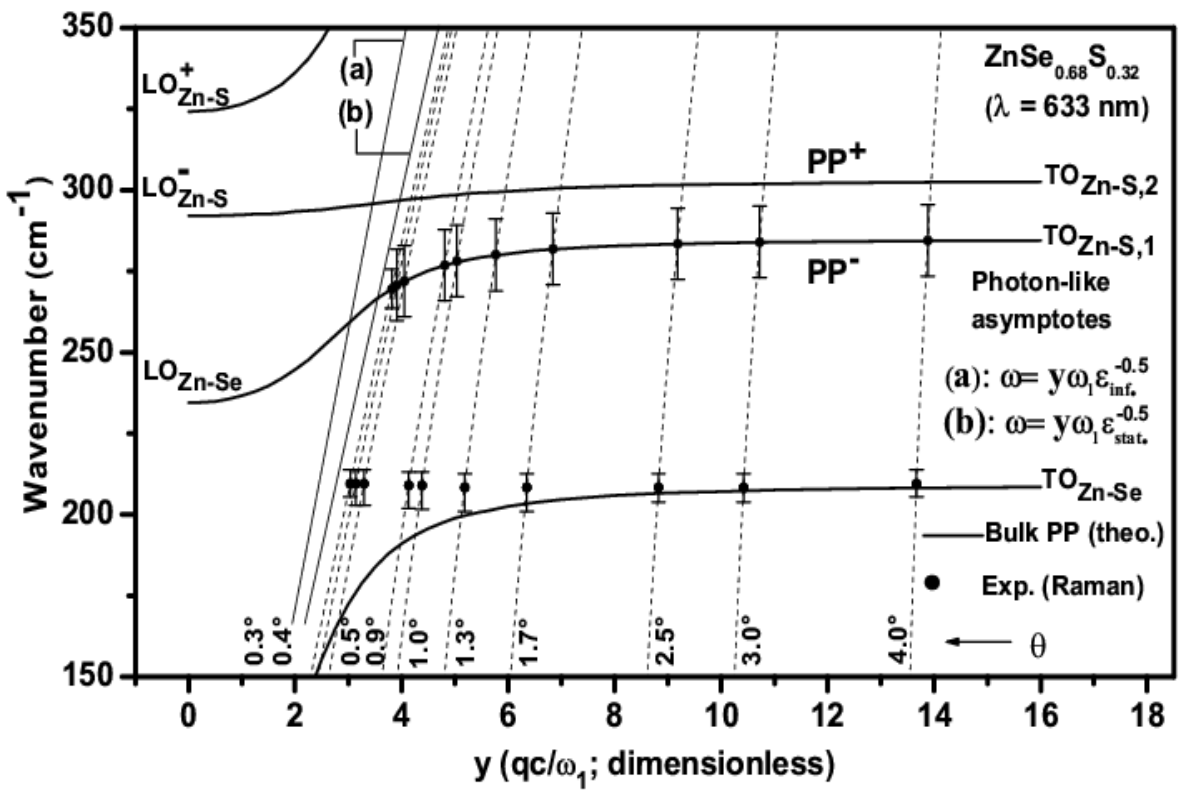

Fig. 3: Comparison between the theoretical and experimental '$\omega$ vs. $q$' multi-*PP* dispersions of $ZnSe_{0.68}S_{0.32}$, as obtained via **Eq. (3)** (thick lines) and by near-forward Raman scattering (symbols with error bars representing the Raman linewidths at half height), respectively. No experimental data is reported for $PP^{+}$ due to a screening at any $\theta$ angle. The relevant '$\theta$ vs. $y$' correspondences per Raman spectrum, in reference to **Eq. (1)**, are added (refer to the oblique-hatched lines), for sake of completeness. The phonon asymptotes in the $q_{\infty}$-regime (*TO* labeling in abbreviation of *PM-TO*'s) and near Γ (*LO* labeling), together with the photon ones (thin lines) far-below (a) and far beyond (b) the phonon resonance are also specified, for reference purpose.

For sake of completeness, we provide in **Fig. 3** a comparison between the theoretical (thin curves) and experimental (symbols) 'ω vs. *q*' multi-*PP* dispersions of $ZnSe_{0.68}S_{0.32}$, as obtained via **Eq. (3)** – in reference to **Fig. 1**, and by near-forward Raman scattering – in reference to **Fig. 2**, respectively.

Summarizing, we perform a near-forward Raman study of the three-mode [1×(*Zn−Se*), 2×(*Zn−S*)] $ZnSe_{0.68}S_{0.32}$ alloy in search of the presumed *LO*-like reinforcement of the (Zn-Se,Zn-S)-mixed $PP^-$ near Γ. The laser excitation as well as the alloy compositions were selected so as to minimize the dispersion of the refractive index, a prerequisite to penetrate deep into the $PP^-$ dispersion in view to address the vicinity of Γ. The *LO*-regime is successfully accessed, as evidenced by the development of $PP^-$ into a giant feature at ultimately small scattering angles, solving the raised issue in the positive sense. The discussion is supported by a contour modeling of the $ZnSe_{0.68}S_{0.32}$ multi-*PP* near-forward Raman lineshapes in their ($q$,$\theta$)-dependence within the formalism of the linear dielectric response.

***

We would like to thank P. Franchetti and J.-P. Decruppe for technical assistance in the Raman measurements, C. Jobart for sample preparation, and A.V. Postnikov for useful discussions and careful reading of the manuscript. This work has been supported by the "Fonds Européens de DEvelopment Régional" of Region Lorraine (FEDER project N°. 34619).